**Quantitative piezoelectric force microscopy: Influence of tip shape, size, and contact geometry on the nanoscale resolution of an antiparallel ferroelectric domain wall**

Lili Tian[1], Aravind Vasudevarao[1], Anna N. Morozovska[2], Eugene A. Eliseev[3], Sergei V. Kalinin[4], Venkatraman Gopalan[1*]

[1] Materials Science and Engineering, Pennsylvania State University, University Park, PA 16802
[2] Institute of Semiconductor Physics, National Academy of Science of Ukraine, 45, pr. Nauki, 03028 Kiev, Ukraine
[3] Institute for Problems of Materials Science, National Academy of Science of Ukraine, 3, Krjijanovskogo, 03142 Kiev, Ukraine
[4] Materials Science and Technology Division and Center for Nanophase Materials Sciences, Oak Ridge National Laboratory, Oak Ridge, TN 37831

*email: vgopalan@psu.edu

**Abstract**
The structure of a single antiparallel ferroelectric domain wall in $LiNbO_3$ is quantitatively mapped by piezoelectric force microscopy (PFM) with calibrated probe geometry. The PFM measurements are performed for 49 probes with the radius varying from 10 to 300 nm. The magnitude and variation of the experimental piezoelectric coefficient across a domain wall matches the profiles calculated from a comprehensive analytical theory, as well as 3-dimensional finite element method simulations. Quantitative agreement between experimental and theoretical profile widths is obtained only when a finite disk-type tip radius that is in true contact with the sample surface is considered, which is in agreement with scanning electron microscopy images of the actual tips after imaging. The magnitude of the piezoelectric coefficient is shown to be independent of the tip radius, and the PFM profile width is linearly proportional to the tip radius. Finally we demonstrate a method to extract any intrinsic material broadening of the ferroelectric wall width. Surprisingly wide wall widths of 20- 200nm are observed.

# I. INTRODUCTION: FERROELECTRIC WALL WIDTH AND COERCIVE FIELDS

The extremely small width of ferroelectric domain walls, typically of the order of 1-2 lattice units,[1, 2] has attracted significant interest in these materials as potential data storage media. The "up" and "down" polarization states in a ferroelectric created by localized electric field serve as data storage bits, with > 10 Tbit/in$^2$ storage density demonstrated recently.[3] Domain shaping on diverse shapes and length scales is also critical to THz surface acoustic wave devices, nonlinear optical frequency conversion, as well as electro-optic steering, dynamic focusing and beam shaping devices. In these applications, domain wall width determines the minimum feature size and maximum operation frequency of the device.[4] Domain wall width also directly influences the dynamics of wall motion[5, 6] A recent intriguing hypothesis[7, 8] suggests that even minute broadening of a domain wall dramatically lowers the coercive fields in ferroelectrics through lowering of the threshold field for wall motion against intrinsic lattice friction. Thus determining the nanoscale structure of a wall is of fundamental interest to the field of ferroelectrics.

To date, the primary means of investigating wall widths on unit cell level has been transmission electron microscopy (TEM).[9, 10, 11, 12, 13] The original TEM studies of a ferroelectric 180° domain wall in a related material, lithium tantalate, concludes that wall width cannot be resolved down to their resolution limit of 0.28nm.[10] Recently, an improved TEM technique has demonstrated that charged 180° walls in lead zirconate titanate thin films can be up to 4-5nm.[14] In parallel, direct imaging of strain at these walls using synchrotron X-ray,[15, 16] index contrast using near-field scanning optical microscopy[17], and excitation emission spectroscopy[18, 19] reveal property changes on length scales of 1-30µm. Thus the scale of 1nm-1µm linking atomic structure of the wall and macroscopic properties of the ferroelectric has been less explored.

In this work, we analyze the structure of a 180° domain wall on the 1-100 nm length scales by scanning probe microscopy with calibrated probe geometry. Previously, atomic force microscopy measurements of surface topography at twinned 90º walls have been used to derive wall widths of ~1.5 nm and determine defect effect on wall broadening.[20, 21] However, there is no intrinsic topography associated with 180º domain walls, necessitating detection of primary order parameter, $P_s$. Piezoelectric Force Microscopy (PFM) detects the surface displacements, $U_i$, related to piezoelectric strain $\varepsilon_{ij}=d_{kij}E_k$ induced by applying an oscillating electric field $E_k$ to the tip in contact with the sample surface.[22, 23] Since piezoelectric coefficients are related to the order parameter components as $d_{ijk}=\gamma_{ijkl}P_l$, where $P_l=P_s$ is the spontaneous polarization, and $\gamma_{ijkl}$ is the electrostrictive coefficient of the material, measurement of the piezoresponse across a wall is expected to provide direct information on the primary order parameter, $P_s$ across a wall. (The fourth-order electrostrictive tensor, $\gamma_{ijkl}$ is not expected to change across the wall, since it is a property of the prototype paraelectric phase and is symmetric with respect to inversion symmetry across the wall). PFM technique has been reviewed in many places, and has been extensively used to study ferroelectric domains.[24, 25, 26, 27, 28] However, to date there has been very little experimental and theoretical investigation of the resolution limits of PFM[29, 30] in order to understand widely varying wall width studies[7, 31, 32], thus the exact limits of domain wall width are still highly debated.

This paper presents a rigorous approach to probe the wall width on the nanoscale through PFM. The measurements are performed using 49 probes with calibrated probe geometry in the 10 nm – 300 nm range. Since the goal is to extract information regarding the intrinsic wall width of an antiparallel wall, the first step is to quantitatively understand the PFM imaging technique, and its resolution limits. By varying the effective radius of the probe, carefully characterizing the details of the tip shape and tip-sample contact region (Section II), and combining it with 3-dimensional Finite Element Modeling (Section III), and analytical theory (Section IV), we demonstrate that the PFM profiles can be quantitatively understood. We demonstrate that intrinsic or extrinsic wall broadening can be extracted by a careful comparison of experiments and theory (Section V).

**II. PFM EXPERIMENTS**
**A. Tip Shape and Contact**

Antiparallel domain walls in congruent lithium niobate are the focus of this study. The point group symmetry of $LiNbO_3$ is $3m$ and the polarization is along $+z$ direction ($+P_z$) or $–z$ direction ($-P_z$). The walls are typically parallel to the crystallographic $y$-$z$ mirror planes. Hence the wall coordinates are defined as $x$ perpendicular to the wall, $y$ along the wall, and $z$- along the polarization direction.

The origin of the contrast in vertical piezoelectric force microscopy arises from piezoelectric deformation due to the converse piezoelectric effect. The application of a localized external electric field to a piezoelectric material, results in a local strain, and consequently displacement of the surface. In a contact AFM mode, the tip is expected to follow this deformation of the surface. The vertical PFM (bending mode) detects the displacement of sample surface perpendicular to the sample surface. The displacement detection sensitivity of ~pm is enabled through the use homodyne detection using lock-in amplifier. For lithium niobate (point group $3m$), the piezoelectric tensor for lithium niobate has 4 independent nonzero coefficients ($d_{31}$, $d_{33}$, $d_{22}$, and $d_{15}$).

Since PFM is a contact mode technique, abrasion of the tip occurs during imaging[33, 34] which changes the tip shape and the field distribution under the tip. The PFM literature typically approximates the tip to be an ideal sphere or disc with a radius $r$. The contact of the tip to the sample is either considered to be an ideal point contact, or more commonly, a dielectric gap of the order of 0.1-1nm is assumed between the tip and the sample. Exact analytical expressions for the field distribution around such tip shapes are well-known.[35, 36] However, below, we show that these assumptions of tip shape and tip-sample contact region are limited.

In order to rigorously describe the tip shape, we imaged the end of each tip using Field Emission Scanning Electron Microscope (FESEM), after scanning a few PFM line scans across a domain wall. These tips can be divided into two sets, 1 and 2, as they are referred to in Figure 1. As seen in Figure 1(a), tip set-2 looks more disk-like, in that its end is flat with a circular contact area of radius $r$. A majority of the tips were of this type. It is modeled as a disk of charge of radius $r$ on the surface of the sample. Tip set-1, seen in Figure 1(b) appears sphere-like, and was usually seen for very small tip radius. The radius of the *contact* area for tip set-2 can be determined by intersecting a straight line (surface of the sample) with the end of the tip in the image; however, a more systematic method that yielded the same results was followed: drawing an imaginary circle at the end of the tip, and taking the cross sectional area of radius *r* at a depth of $h\sim$1-2 nm (1-2 *c* lattice units) depth from the tangent to the surface of the circle. The *h* should however, not be construed as a real indentation, but is rather chosen phenomenologically as an engineering parameter that accounts for the SEM image diffuseness at the end of the imaged tip. This is used to estimate the contact area formed during the wear process. We find that this uncertainty in *h* (1-2 nm) and hence in *r* does not affect our conclusions, which are dominated by experiments with tip set-2. In particular, for $r\sim$10 nm, the observed PFM domain wall resolution is ~100 nm, well beyond broadening anticipated from conventional indentation models (e.g. Hertzian).

**B. PFM profiles across a wall**

The PFM signal is typically a complex displacement, $\tilde{U} = U_R + iU_I = U_o e^{i\theta}$. The phase, $\theta$, refers to the relative phase of the tip displacement with respect to the phase of the alternating voltage applied to the PFM tip. The pure electromechanical signal $\tilde{U}$ can be clearly distinguished in the complex plane from any background signal as described in literature.[37] This background subtraction described in Ref. 37 is critical for quantitative analysis of the PFM profiles, and performing it eliminates the frequency dependence of the PFM signal in the frequency range of 20-100kHz used in this study. A monotonic frequency dispersion of the pure PFM signal still remains below the frequency of 20kHz presumably due to lock-in non-idealities, and hence this frequency range is excluded from this study. Figure 2 depicts example profiles of the $U_R$ and $U_I$ as a function of the wall normal coordinate *x*. Note that with appropriate background subtraction, the pure electromechanical displacement is entirely along the real part $U_R = U_o \cos\theta$, (Figure 2(a)) while $U_I = U_o \sin\theta$ (Figure 2(b)) is zero. Since the displacement component normal to the surface ($U_z$) was measured (vertical PFM), $U_R \sim \pm U_o = \pm U_z$. This indicates that the phase $\theta$ is 0 or $\pi$. The amplitude $|U_z|$ and phase, $\theta$ are shown in Figure 2(c) and Figure 2 (d), respectively. Such experimental wall profiles were measured for a series of tip radii, which will be discussed further in relation to theory and simulation in Section V.

**C. Tip size dependence of the PFM amplitude and width**

Experimentally, there are two important experimental parameters that were extracted from these profiles: amplitude, $|U_z|$, and width, $\omega_{PFM}$. The PFM amplitude was calibrated using a poled PZT ceramic sample uniformly electroded on both sides, whose piezoelectric coefficient in pm/V was independently measured using a piezometer, which applies a stress and measures the open loop potential generated across the material. The surface displacement of this PZT sample in response to an applied voltage from a PFM tip was then used to calibrate the PFM signal. This calibration in turn was used to quantify the piezoelectric coefficient of the LiNbO$_3$ sample. The calibrated amplitude of the PFM response away from the wall in units of $d_{eff}$ (pm/V) is plotted as a function of the experimentally determined tip radius *r*, in Figure 3. An important conclusion seen from Figure 3 is that the $d_{eff}$ is independent of the tip radius used. Remarkably, thus calibrated magnitude of the $d_{eff}$ in pm/V as well as its invariance with respect to tip radius also agrees with both the analytical theory [38, 30] and numerical simulations described next in Sections III and IV.

Figure 4 shows the width $2\omega_{PFM}$ of the PFM response across a single 180° domain wall as a function of the experimentally determined tip radius $r$. The $\omega_{PFM}$ refers to the half width where the PFM response reaches ±0.76 of the saturation PFM value away from the center of the wall. This saturation value was taken as the PFM value at ~±1.8μm from the center of the wall. The PFM width decreases linearly with the tip radius, except at the smallest tip radii, where deviations from linearity are observed. The relationship between these deviations and the intrinsic wall width is discussed in Section V. Two modeling approaches were employed, namely, analytical theory and Finite Element Modeling (FEM) as described below.

### III. FINITE ELEMENT SIMULATION OF THE PFM RESPONSE

In order to understand the PFM response quantitatively, we also perform Finite Element Method (FEM) modeling of the imaging process using the commercial ANSYS program as well as analytical theory using the decoupled approximation.[27] The FEM approach described in ref. 27 in detail, includes a complete description of the geometry of the tip, the sample, and the contact region, numerical calculation of the electric field distribution with a constant potential applied to the tip, and using the computed potential distribution on the sample surface as boundary condition to calculate the piezoelectric deformation of the surface (Figure 5(a)). Input to the FEM includes the complete dielectric tensor, elastic tensor and piezoelectric tensor of the sample, thus providing a rigorous 3-D approach. (The $d_{22}$ coefficient ignored in analytical theory was included in FEM, and its effect on the PFM was shown to be minimal as well and it is confirmed later that this is a valid assumption.). A single domain wall, parallel to one of the three degenerate *y-z* crystal physics planes was defined in the simulation by flipping the crystal physics axes, *y* and *z* across the wall. Figure 5(b) depicts the surface deformations simulated by FEM for three different locations of the tip across a single *step-like* 180º domain wall. (Three separate simulations for the three tip positions have been merged in this plot). Using a series of positions of tip across the wall, continuous FEM line profiles of $U_z$ were generated.(see video clip; online supplementary information) All simulations with FEM were performed only with step-like wall, and did not include any intrinsic broadening or diffuseness, since it was not numerically feasible in the software used. Diffuseness is included in the analytical theory, described later.

We explore two different models for the tip-sample surface interaction: (a) the sphere-plane model, and (b) the disc-plane model, where the sphere or the disc refers the tip shape and the plane refers to the sample surface. In both models, there are three electrostatic boundary conditions to satisfy:
1) An equipotential AFM tip surface equal the electrical potential, *V*, applied to the tip.
2) The tangential component of electric field ***E*** is continuous across the interface between the dielectric medium (air) and the dielectric specimen;
3) The normal component of electric displacement ***D*** is continuous across the interface between the dielectric medium and dielectric specimen.

The conical part the AFM tip contribution in a spherical tip can be modeled using the line charge model developed by Huang *et. al*.[39] The conical part of AFM tip contribution in a disc tip was modeled using ANSYS. The conical part was simplified as a 20μm-long metallic coated silicon with a full cone angle of 30°. By applying the calculated potential as the boundary condition for the piezoelectric coupling simulation using ANSYS, the deformation of the ferroelectric resulting from the AFM tip can be simulated. Piezoelectric response of a single domain lithium niobate *in true contact* with the tip was simulated with finer meshing of the sample surface below the tip and coarser meshing away from the tip. The meshing was adjusted until the results converged and became independent of the fineness of the meshing system.

## IV. ANALYTICAL THEORY OF THE PFM RESPONSE ACROSS A DIFFUSED DOMAIN WALL

**A. Resolution function approach**

Since the goal of this work is the extraction of a 180° domain wall diffuseness, we consider a single domain wall with the strain piezoelectric coefficient tensor terms $d_{klj}$ dependent only on lateral coordinates *x* and *y* perpendicular to the polarization direction. The system is considered uniform in the polarization direction, *z*. In LiNbO$_3$ in particular, the domain walls tend to be parallel to the crystallographic *y-z* planes, hence the spatial dependence needs to be considered only as a function of the wall normal coordinate, *x*. The surface displacement vector $U_i(z)$ (measured PFM piezoresponse) is given by the convolution of piezoelectric tensor coefficients $d_{klj}(x)$ with the resolution function components $W_{ijkl}(x,y)$ as proposed in Ref. 30. Since in many cases, the inhomogeneous distribution of piezoelectric coefficients are similar, e.g. for ferroelectrics they are determined by the polarization distribution, hereafter we denote the inhomogeneous part of the piezoelectric coefficients as function $\beta(x)$. In this approximation, the components of the surface displacement below the tip can be written as follows[30]

$$U_i(x) = \int_{-\infty}^{\infty} dx' \int_{-\infty}^{\infty} dy\, W_{ijkl}(-x',-y) d_{lkj}^{bulk} \beta(x-x'), \qquad (1)$$

Here $d_{lkj}^{bulk}$ are constant piezoelectric coefficients of bulk material. The resolution function is introduced as:

$$W_{ijkl}(x,y) = c_{kjmn} \int_0^{\infty} dz \frac{\partial G_{im}(-x,-y,z)}{\partial x_n} E_l(x,y,z). \qquad (2)$$

Here $E_l$ is the component of the external electric field produced by the probe, $c_{kjmn}$ are stiffness tensor components, $\partial G_{im}/\partial x_n$ is a semi-space elastic Green tensor derivatives on Cartesian coordinate $x_n = \{x,y,z\}$. For most inorganic ferroelectrics, the elastic properties are weakly dependent on orientation and hereinafter the material can be approximated as elastically isotropic. Corresponding Green's tensor $G_{ij}(x,y,z)$ for elastically isotropic half-plane is given by Lur'e[40] and Landau and Lifshitz[41].

Using decoupling approximation,[42, 27] and resolution function approach[29] for transversally isotropic media,[30] vertical piezoelectric response of isolated 180°-domain wall in the inhomogeneous electric field of the probe tip has the form:[43]

$$d_{33}^{eff}(x) = \frac{U_3(x)}{V} = \frac{1}{2V} \int_{-\infty}^{\infty} U_3^{step}(x-x') \frac{\partial \beta(x')}{\partial x'} dx'. \quad (3)$$

Here $V$ is electric bias applied to the probe tip; $U_3(x)$ is the surface displacement below the tip located at distance $x$ from the plain domain wall located at $x = a_0$. The surface displacement $U_3^{step}(x)$ of a *step*-like infinitely thin domain wall is derived in Ref. [30]. Below we list the final close-form expression:

$$U_3^{step}(x) = V\left(d_{31}\left(\frac{x(1+\nu)f_{313}}{|x|+C_{313}z_o} - \frac{xf_{333}}{|x|+C_{333}z_o}\right) + \frac{xf_{333}d_{33}}{|x|+C_{333}z_o} + \frac{xf_{351}d_{15}}{|x|+C_{351}z_o}\right). \quad (4)$$

Here $d_{lm} \equiv d_{lkj}^{bulk} = d_{ljk}^{bulk}$ in Vogt notation, $\nu$ is the Poisson ratio. Characteristic distance $z_o$ is determined by the parameters of the tip. In the effective point charge model it is the charge-surface separation. If we approximate the tip by the metallic disk of radius $r$ in contact with surface, then $z_o = 2r/\pi$. The expressions for material anisotropy constants $f_{ijk}$ and $C_{ijk}$ are given in Appendix A.

Using (4), we can derive a simple approximation for the effective width, $\omega_{PFM}$ of infinitely thin domain wall (measured as distance between the points where the response is equal to $\pm(1-\eta)$ fraction of saturation polarization:

$$\omega_{PFM} \approx 2z_o \frac{1-\eta}{\eta} \frac{((1+\nu)f_{313}C_{313} - f_{333}C_{333})d_{31} + f_{333}C_{333}d_{33} + f_{351}C_{351}d_{15}}{((1+\nu)f_{313} - f_{333})d_{31} + f_{333}d_{33} + f_{351}d_{15}} \quad (5)$$

It should be noted that we have neglected the contribution of $d_{22}$ and related terms, since their contribution far from the wall is exactly zero in the framework of the decoupling approximation model.[44]

In the effective point charge approximation of the tip, electric field and dielectric anisotropy $\gamma \approx 1$, vertical piezoresponse $d_{33}^{eff}$ at a distance $x$ from the exponential domain wall profile $\beta(x, \omega_0) = (1 - \exp(-|x|/\omega_0))\text{sign}(x)$ located at $x = 0$ admits closed-form analytical representation:

$$d_{33}^{eff}(x, \omega_0, z_0) = -\left(\left(\frac{1}{4} + \nu\right)d_{31} + \frac{3}{4}d_{33} + \frac{d_{15}}{4}\right)\beta(x, \omega_0) -$$

$$-\left(\left(\frac{1}{4} + \nu\right)d_{31} + \frac{3}{4}d_{33}\right)\frac{z_0}{8\omega_0}\left(\exp\left(-\frac{|x|}{\omega_0}\right)F\left(\frac{z_0}{4\omega_0}\right) - F\left(\frac{|x|}{\omega_0} + \frac{z_0}{4\omega_0}\right)\right)\text{sign}(x) - \quad (6)$$

$$-\frac{d_{15}}{4}\frac{3z_0}{8\omega_0}\left(\exp\left(-\frac{|x|}{\omega_0}\right)F\left(\frac{3z_0}{4\omega_0}\right) - F\left(\frac{|x|}{\omega_0} + \frac{3z_0}{4\omega_0}\right)\right)\text{sign}(x)$$

Here $\omega_0$ is the wall $\beta(x, \omega_0)$ intrinsic width, function $F(x) = \exp(-x)\text{Ei}(x) - \exp(x)\text{Ei}(-x)$, where $\text{Ei}(x) = \int_{-x}^{\infty} dt \exp(-t)/t$ is the tabulated exponential integral function. An approximation $F(x) \approx \frac{2x}{x^2+c} - \frac{2x}{x^2+c}\ln\left(\frac{|x|}{|x|+1}\right)$ (constant $c = \frac{1}{1-EulerGamma} \approx 2.365$) is valid with 3% accuracy for all $x$-range. The

first term in Eq.(6) is the ideal image $\beta(x,\omega_0)$ of domain wall intrinsic profile. Near and far from the wall plane, the following expansions are valid:

$$d_{33}^{eff}(x) \approx \begin{cases} -\frac{x}{\omega_0}\left( \begin{array}{l} \left(\left(\frac{1}{4}+\nu\right)d_{31}+\frac{3}{4}d_{33}\right)\left(1+\frac{z_0}{4\omega_0}\exp\left(\frac{z_0}{4\omega_0}\right)\text{Ei}\left(-\frac{z_0}{4\omega_0}\right)\right)+ \\ +\frac{d_{15}}{4}\left(1+\frac{3z_0}{4\omega_0}\exp\left(\frac{3z_0}{4\omega_0}\right)\text{Ei}\left(-\frac{3z_0}{4\omega_0}\right)\right) \end{array} \right), & |x|<\omega_0 \\ \text{sign}(x)\left( \begin{array}{l} -\left(\left(\frac{1}{4}+\nu\right)d_{31}+\frac{3}{4}d_{33}+\frac{d_{15}}{4}\right)+\frac{d_{15}}{4}\frac{3z_0}{4|x|+3z_0}+ \\ +\left(\left(\frac{1}{4}+\nu\right)d_{31}+\frac{3}{4}d_{33}\right)\frac{z_0}{4|x|+z_0} \end{array} \right), & |x|\gg\omega_0 \end{cases}$$ (7)

**B. Contribution of the conical part to the disk model for the tip**

It is known that the conical part of the probe, as well as the tip-surface contact area contributions to the electrostatic potential broaden and diffuse the piezoresponse profile of the wall. To estimate the cone effects in PFM imaging, the conical part was modeled by a line charge,[39] and the contact area by a disk touching the sample surface, as proposed elsewhere.[39]

Using electric field superposition principle, below we consider the probe electrostatic potential $\varphi(\rho,z)$ as the sum of effective line charge potential, $\varphi_L$, point charge potential $\varphi_q$ and disk potential $\varphi_D$:

$$\varphi(\rho,z) = \varphi_L(\rho,z) + \varphi_q(\rho,z) + \varphi_D(\rho,z),$$ (8)

Here the radius $\rho = \sqrt{x^2+y^2}$. Normalization in Eq. (7) is such that $\varphi_L(0,0)+\varphi_D(0,0)+\varphi_q(0,0) \approx V$, the applied potential to the tip. The conical part potential $\varphi_L$ is modeled by the linear charge of length $L$ with a constant charge density $\lambda_L = 4\pi\varepsilon_0 V \Big/ \ln\left(\frac{1+\cos\theta}{1-\cos\theta}\right)$, where $\theta$ is the cone apex angle. Additional point charge potential $\varphi_q$ is chosen to reproduce the conductive tip surface as closely as possible by the isopotential surface $\varphi(\rho,z)=V$. The contact area potential is modeled by a disk of radius, $r$ (see Appendix B). Numerical calculations proved that the charge $q$ is located at the end of the line at a distance of approximately the disk radius $r$ from the surface, and that $q \approx 4\pi\varepsilon_0 Vr$ for a wide range of cone angles $\theta$. It is clear from the Figure 6 that for a chosen geometry, the isopotential surface $\varphi(\rho,z)=V$ reproduces the conductive tip shape in the vicinity of the surface for a wide range of cone angles $\theta$. Next we calculate domain wall profiles including different parts of the probe.

**C. Diffused domain wall profile**

Analytical theory predictions of the vertical PFM response near the single domain wall in LiNbO$_3$ are shown in Figure 7. The influence of the tip radius itself on the wall profile is shown in for a step-like wall (diffuseness $\omega_o=0$), clearly indicating the PFM wall width increase with radius $r$. The influence of domain wall diffuseness is shown in where the piezoelectric coefficient profile is $d_{lkj}(x) = d_{lkj}^{bulk} \tanh(x/\omega_0)$ (i.e. the intrinsic profile $\beta(x) = \tanh(x/\omega_0)$). This is chosen to mimic the polarization variation, $P_3$ across a 180°-domain wall, given by $P_3(x) = P_3^{bulk} \tanh(x/\omega_0)$, since the piezoelectric coefficients and the polarization are linearly related by the electrostriction tensor. As expected, domain wall diffuseness broadens the PFM wall profile for a give tip radius $r$. Similar results can be obtained for an exponential wall diffuseness profile by using Eqs. (6) and (7).

The combination of both a change in tip radius, $r$ and a change in the wall diffuseness $\omega_o$ was previously shown in a series of theory plots in Figure 4, along with experimental data points. The PFM wall width increases linearly with the tip radius $r$ for a step-like wall. Domain wall diffuseness, $\omega_o$ adds significant nonlinearity to these curves for approximately tip radii $r<\omega_o$. The general *quantitative* agreement between the theory and experiments is excellent. It also suggests that there may be significant domain wall diffuseness in LiNbO$_3$ crystals. This issue is discussed in greater detail in Section V.

## IV. COMPARISON BETWEEN EXPERIMENTS, SIMULATION, AND THEORY
### A. PFM amplitude and width

Figure 3 (shown earlier) plots the maximum effective piezoelectric coefficient, $d_{eff}=U_z/V$ measured and simulated away from the wall, where $V$ is the maximum voltage on the sample surface. The experiments, analytical theory and FEM simulations show excellent *quantitative* agreement with each other. Within error bars, both experimental and simulated values of the maximum displacement $U_z$ and the corresponding $d_{33}^{eff}$ are relatively insensitive to the tip radius, and the presence of the cone. These agreements provide us confidence in our experimental and simulation methods.

Figure 8 shows a comparison of the $\omega_{PFM}$ as a function of tip radius $r$ for various tip models. It is clear from the comparisons that spherical tips with a point contact do not agree quantitatively with experimental results. The predicted PFM widths in this case are much smaller than those experimentally observed. Introducing an imaginary dielectric layer gap (air) of 2 nm (which is large) between the spherical tip and the surface increases the predicted $\omega_{PFM}$, but still is considerably less than the experimentally measured widths. Only the disk-type tip model, including the conical section shows the best agreement with the experiments. In the context of the presented work, these results rule out the spherical tip models and the possibility of dielectric gaps. True contact with a disk type tip and no dielectric gap therefore reflects the true nature of the PFM imaging presented here.

### B. PFM wall widths versus Intrinsic wall width

Finally, we pose the primary question we began with: What information can we extract about the intrinsic wall width of a ferroelectric using PFM? This question is naturally related with the question of what the resolution of the PFM technique is. Finite Element Method and analytical theory show that for a point contact of the tip, and in the absence of the conical part of the tip, the PFM wall width tends to zero. In other words, infinite resolution is, in principle, possible. However, this is not practical due to the presence of the cone. With the cone part of the tip, (20μm-long metallic coated silicon with a full cone angle of 30°), the linear extrapolation of the FEM predicted PFM width to zero tip contact radius is ~11±10 nm, which is statistically zero width. However, practical consideration of a finite tip contact radius, and abrasion of the tip while in contact with the sample results in experimentally measured PFM widths on the order of ~100 nm.

Nonetheless, if enough statistical data points are collected as in Figure 4, one can begin to make some conclusions about intrinsic ferroelectric wall widths. Experimental results in Figure 4 show a significant scatter in PFM widths up to 100nm and more particularly in the small tip contact radius range. Does this reflect information regarding the intrinsic ferroelectric wall width at the surface?

To answer this, we first explore extrinsic factors such as tilted domain walls with respect to the surface normal. Cross-sectional polishing, etching and imaging of domain walls reveal that within error of measurement, the domain walls in our samples were typically <0.1°-0.5°, and in some extreme cases, as much as 2.6° away from the surface normal. FEM simulations with tilted walls were performed, with no cone, and zero intrinsic domain wall width. The PFM wall width is apparently broadened by this tilt. For a 5° tilt, the extrapolated PFM wall width at zero tip contact radius was ~10nm. For typical values of 0.1° tilt, the corresponding PFM width at zero tip contact radius was simulated to be <0.5nm. Thus, one can reasonably say that domain wall tilts cannot account for full range of scatter of PFM widths (at zero tip contact radius) over 10nm.

Finally, one can incorporate an intrinsic wall diffuseness, $2\omega_o$, in the analytical theory in terms of piezoelectric tensor distribution $d_{lkj}(x) = d_{lkj}^{bulk} \tanh(x/\omega_0)$. (This was not possible with the ANSYS software, hence FEM simulation was not performed). Figure 4 shows the theoretical PFM width predictions for different values of intrinsic domain wall diffuseness. One notices that the PFM wall width versus tip contact radius becomes nonlinear for small tip radius, and this nonlinearity increases for larger intrinsic wall diffuseness. The scatter of experimental data points fall below the $\omega_o$ ~100nm theory curve, suggesting a range of intrinsic wall diffuseness at the surface of congruent LiNbO$_3$. Figure 9(a) shows that for large tip radius ($r$=200nm), the match between experiments, FEM and analytical theory is excellent, and $\omega_o$ has little influence on the wall profile. However, as the tip radius $r$ approaches $\omega_o$, the simulation of a sharp (step – like) domain wall *does not* faithfully match the experimental line profile. Figure 9(b) shows that an excellent fit is obtained only when $2\omega_o$ =60nm for that particular line profile. This analysis was repeated for all the data points in Figure 4, and $2\omega_o$ is found to vary from 20-200nm.

Further confidence in this assertion arises from independent measurements of domain wall width in congruent LiNbO$_3$ using scanning nonlinear dielectric microscopy (SNDM) technique. Cho et al. have recently demonstrated imaging of the Si(111)7x7 surface[45,46] atomic structure using similar 2$^{nd}$ and 3$^{rd}$ order capacitance terms in SNDM showing <0.5nm resolution of SNDM. Surface domain wall widths of 20-150nm have been measured in LiNbO$_3$ and LiTaO$_3$ crystals by this technique.[47] Images in Ref. 47 reveal that the larger wall widths arise when large polar and dielectric defects exist adjacent to the wall. While SNDM has the resolution to reveal these defects, the PFM technique detects it as an effective broadening of the domain wall. Since PFM detects a third rank polar property, and SNDM images show dielectric defects using third rank dielectric tensor, these wall broadenings can be concluded to be polar in nature, and reflecting polarization fluctuations. Similar SNDM studies in isostructural lithium tantalate,[48] performed in cross-sectional, *y*-cut geometry of the crystal reveals that the wall width decreases from ~15nm just below the surface to ~2.5-1nm at a depth of ~50-100nm from the z-surface of the crystal. This shows that surfaces also broaden domain walls. Thus, we conclude that dielectric/polar defects and surfaces can broaden antiparallel ferroelectric domain walls, and this broadening, on the scale of tens of nanometers is being detected in these PFM studies.

## V. CONCLUSIONS

In conclusion, we have shown that piezoelectric force microscopy (PFM) can be a quantitative tool for probing piezoelectric materials, and particularly ferroelectric domains and domain walls. A PFM tip with finite contact area and in true contact with the sample surface gives the best agreement between PFM experiments, analytical theory, and finite element modeling of the PFM response across a domain wall. The PFM amplitude is independent of the tip radius. The PFM width across a sharp domain wall is linear with tip contact radius, and is predicted in theory to provide infinite resolution for a point contact. However, practically, the presence of conical part of the tip, any slight tilts of the wall with respect to surface normal, and importantly, the abrasion of the tip on contact with the sample leads to finite resolution on the order of ~10-20nm. Theory predicts that walls with finite intrinsic diffuseness will lead to a non-linear relationship between PFM width and tip contact radius, particularly for small tip radii on the scale of wall diffuseness. Using a combination of PFM experimental line profiles across 180° domain walls, analytical theory, and FEM simulations, we conclude that real domain walls on the z-surface of lithium niobate show broadened walls on the scale of *$2\omega_b$*~ 20-200nm, with considerable scatter from location to location. The scatter arises from the surface influence, as well as the presence of dielectric and polar defects adjacent to walls, which have been imaged by SNDM technique and reported in literature in these materials.[47, 48, 49] The PFM results show that these defect-domain wall interactions lead to an effective broadening of the walls in terms of polarization.

VG wishes to gratefully acknowledge financial support from the National Science Foundation grant numbers DMR-0602986, 0512165, 0507146, and 0213623, and CNMS at Oak Ridge National Laboratory. Research was also sponsored in part (SVK) by the Center for Nanophase Materials Sciences, Office of Basic Energy Sciences, U.S. Department of Energy with Oak Ridge National Laboratory, managed and operated by UT-Battelle, LLC.

**APPENDIX A**

More rigorous consideration of disk part[43] leads to the substitution $\frac{x f_{3ij}}{|x| + C_{3ij} z_o} \to \frac{2}{\pi} f_{3ij} \arctan\left(\frac{x}{C_{3ij} z_o}\right)$ in Eq.(4), valid for sharp domain walls.

Constants $f_{ijk}$ and $C_{ijk}$ depend only on the material dielectric anisotropy factor $\gamma = \sqrt{\varepsilon_{33}/\varepsilon_{11}}$ as:

$$f_{351}(\gamma) = -\frac{\gamma^2}{(1+\gamma)^2}, \quad f_{333}(\gamma) = -\frac{1+2\gamma}{(1+\gamma)^2}, \quad f_{313}(\gamma) = -\frac{2}{1+\gamma}. \tag{A.1a}$$

$$C_{313}(\gamma) = \frac{1+\gamma}{8\gamma^2} \, {}_2F_1\left(\frac{3}{2}, \frac{3}{2}; 3; 1 - \frac{1}{\gamma^2}\right), \tag{A.1b}$$

$$C_{333}(\gamma) = \frac{3(1+\gamma)^2}{16\gamma^2(1+2\gamma)} \, {}_2F_1\left(\frac{3}{2}, \frac{5}{2}; 4; 1 - \frac{1}{\gamma^2}\right), \tag{A.1c}$$

$$C_{351}(\gamma) = \frac{3(1+\gamma)^2}{16\gamma^2} \, {}_2F_1\left(\frac{3}{2}, \frac{3}{2}; 4; 1 - \frac{1}{\gamma^2}\right), \tag{A.1d}$$

Here ${}_2F_1(p,q;r;s)$ is the hypergeometric function. In particular case $\gamma=1$, $C_{313}(1) = \frac{1}{4}$, $C_{333}(1) = \frac{1}{4}$, $C_{351}(1) = \frac{3}{4}$, $f_{351}(1) = -\frac{1}{4}$, $f_{333}(1) = -\frac{3}{4}$, $f_{313}(1) = -1$. The relevant constants for LiNbO$_3$ and LiTaO$_3$ are given in Tables 1, 2 below:

**Table A.1: Relevant physical properties of LiNbO$_3$ and LiTaO$_3$**

|         | $\varepsilon_{11}$ | $\varepsilon_{33}$ | $\gamma$ | $d_{15}$ (pm/V) | $d_{31}$ (pm/V) | $d_{33}$ (pm/V) |
|---------|------|------|------|------|------|------|
| LiNbO$_3$ | 85 | 29 | 0.58 | 68 | -1 | 6 |
| LiTaO$_3$ | 54 | 44 | 0.90 | 26 | -2 | 8 |

**Table A.2: Relevant PFM parameters in Eqs.A.1a-d for LiNbO$_3$ and LiTaO$_3$**

|         | $f_{313}$ | $f_{333}$ | $f_{351}$ | $C_{313}$ | $C_{333}$ | $C_{351}$ |
|---------|-------|-------|-------|-------|-------|-------|
| LiNbO$_3$ | -1.26 | -0.86 | -0.14 | 0.24 | 0.21 | 0.68 |
| LiTaO$_3$ | -1.05 | -0.775 | -0.225 | 0.25 | 0.24 | 0.74 |

**APPENDIX B**

We consider the probe electrostatic potential $\varphi(\rho,z)$ as the sum of effective line charge potential, $\varphi_L$, point charge potential $\varphi_q$ and disk potential $\varphi_D$:

$$\varphi(\rho,z) = \varphi_L(\rho,z) + \varphi_q(\rho,z) + \varphi_D(\rho,z), \tag{B.1}$$

Here $\rho = \sqrt{x^2 + y^2}$ is the radial coordinate. Under the condition $\varepsilon_{11,33} \gg 1$ typically valid for the majority of ferroelectrics in air, the following expressions for potential structure are valid:

$$\varphi_L(\rho,z) = \frac{V}{\ln\left(\frac{1+\cos\theta}{1-\cos\theta}\right)} \begin{cases} \frac{2}{1+\kappa} \ln\left(\frac{L+\Delta L + z + \sqrt{(L+\Delta L+z)^2+\rho^2}}{\Delta L + z + \sqrt{(\Delta L+z)^2+\rho^2}}\right), & z > 0 \\ \ln\left(\frac{L+\Delta L + z + \sqrt{(L+\Delta L+z)^2+\rho^2}}{\Delta L + z + \sqrt{(\Delta L+z)^2+\rho^2}}\right) + \\ + \frac{1-\kappa}{1+\kappa} \ln\left(\frac{L+\Delta L - z + \sqrt{(L+\Delta L-z)^2+\rho^2}}{\Delta L - z + \sqrt{(\Delta L-z)^2+\rho^2}}\right), & z < 0 \end{cases} \quad \text{(B.2a)}$$

$$\varphi_q(\rho,z) = \frac{q}{4\pi\varepsilon_0} \begin{cases} \frac{2}{1+\kappa} \frac{1}{\sqrt{(\Delta L + z)^2 + \rho^2}}, & z > 0 \\ \frac{1}{\sqrt{(\Delta L + z)^2 + \rho^2}} + \frac{1-\kappa}{1+\kappa} \frac{1}{\sqrt{(\Delta L - z)^2 + \rho^2}}, & z < 0 \end{cases} \quad \text{(B.2b)}$$

$$\varphi_D(\rho,z) = \frac{2V}{\pi} \arcsin\left(\frac{2r}{\sqrt{(\rho-r)^2+z^2} + \sqrt{(\rho+r)^2+z^2}}\right) \times $$
$$\times \left(1 - \frac{2}{1+\kappa}\left(\frac{\ln(1+L/\Delta L)}{\ln((1+\cos\theta)/(1-\cos\theta))} + \frac{q}{4\pi\varepsilon_0 V \Delta L}\right)\right) \quad \text{(B.3c)}$$

Hereinafter, $\kappa = \sqrt{\varepsilon_{33}\varepsilon_{11}}$ is the effective dielectric constant and $\gamma = \sqrt{\varepsilon_{33}/\varepsilon_{11}}$ is the dielectric anisotropy factor. The conical part potential (B.2a) is modeled by the linear charge of length $L$ with a constant charge density $\lambda_L = 4\pi\varepsilon_0 V / \ln\left(\frac{1+\cos\theta}{1-\cos\theta}\right)$, where $\theta$ is the cone apex angle. Additional point charge potential (B.2b) is chosen to reproduce the conductive tip surface as closely as possible by the isopotential surface $\varphi(\rho,z) = V$. Our numerical calculations show that the charge $q$ is located at the end of the line, at that the distance $\Delta L$ from the surface is approximately equal to the disk radius $r$ and $q \approx 4\pi\varepsilon_0 V \Delta L$ for a typical range of cone angles $\theta$. Contact area potential (B.3c) is modeled by the charged disk of radius $r$.

Using superposition (B.1) we calculate the conical part of the probe electric field $E_L = -\partial\varphi_L/\partial x_l$ and then substitute it directly into Eq.(2).

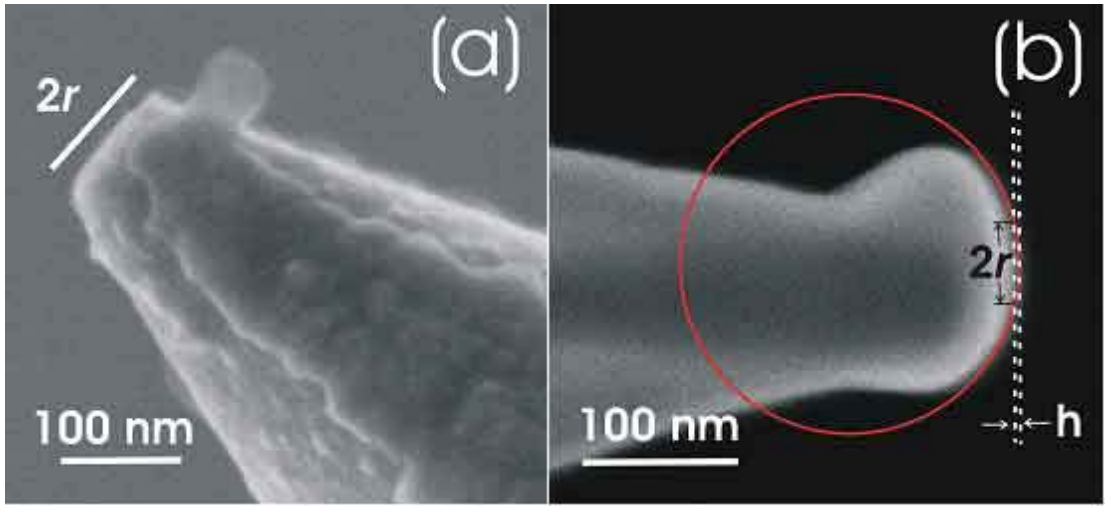

**Figure 1** (color online): (a) Field emission scanning electron microscope image of a used PFM tip from tip set 2, with a circular disk-like end with a radius $r$. (b) Field emission scanning electron microscope image of a used sphere-like PFM tip from tip set 1. The radius $r$ of the contact circle for a weak indentation ($h$~1-2nm or 1 unit cell depth) is used to characterize the radius $r$ of the tip as shown. ($h$ is not to scale in the figure).

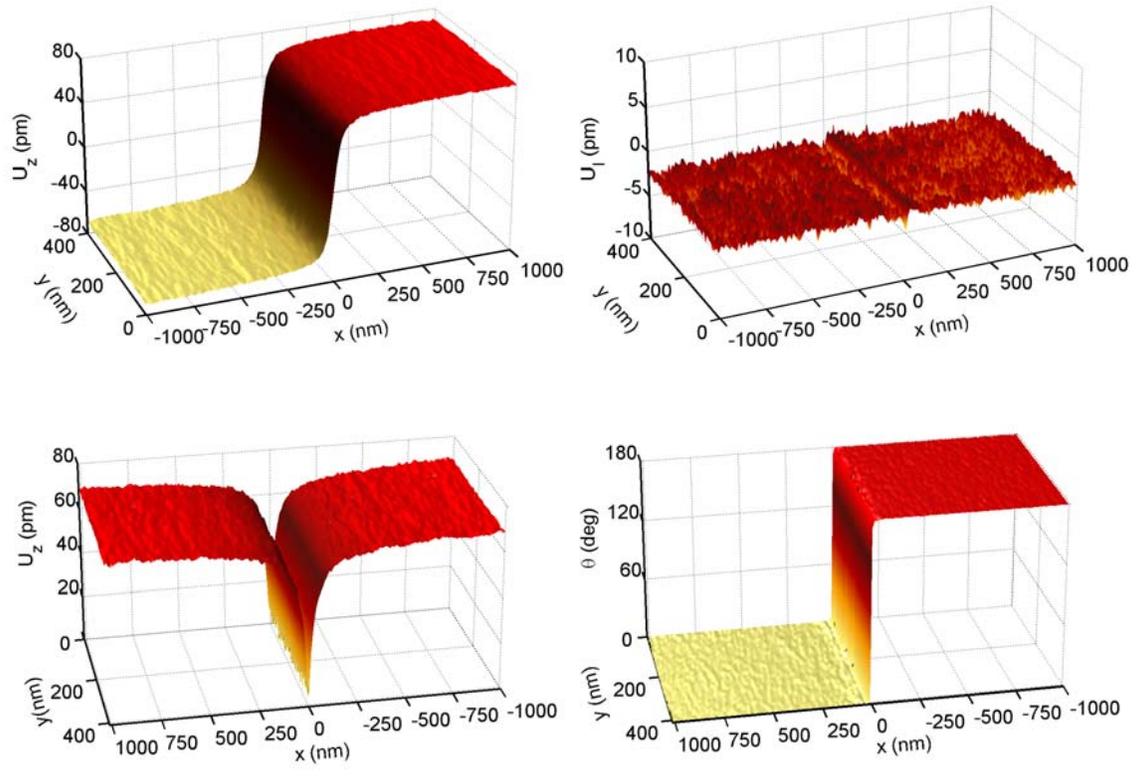

**Figure 2** (color online)*:* An example of complex PFM displacement, $\tilde{U}$ as a function of wall normal coordinate, *x* across a 180° domain wall in lithium niobate. (a) $U_o \cos\theta \sim U_z$, (b) $U_o \sin\theta$, (c) $|U_z|$, and (d) $\theta$. Measurements were made with a Ti/Pt coated Si tip with a tip disk radius of ~50-60 nm. An oscillating voltage of 5Vrms, at 42.35kHz was applied to the tip.

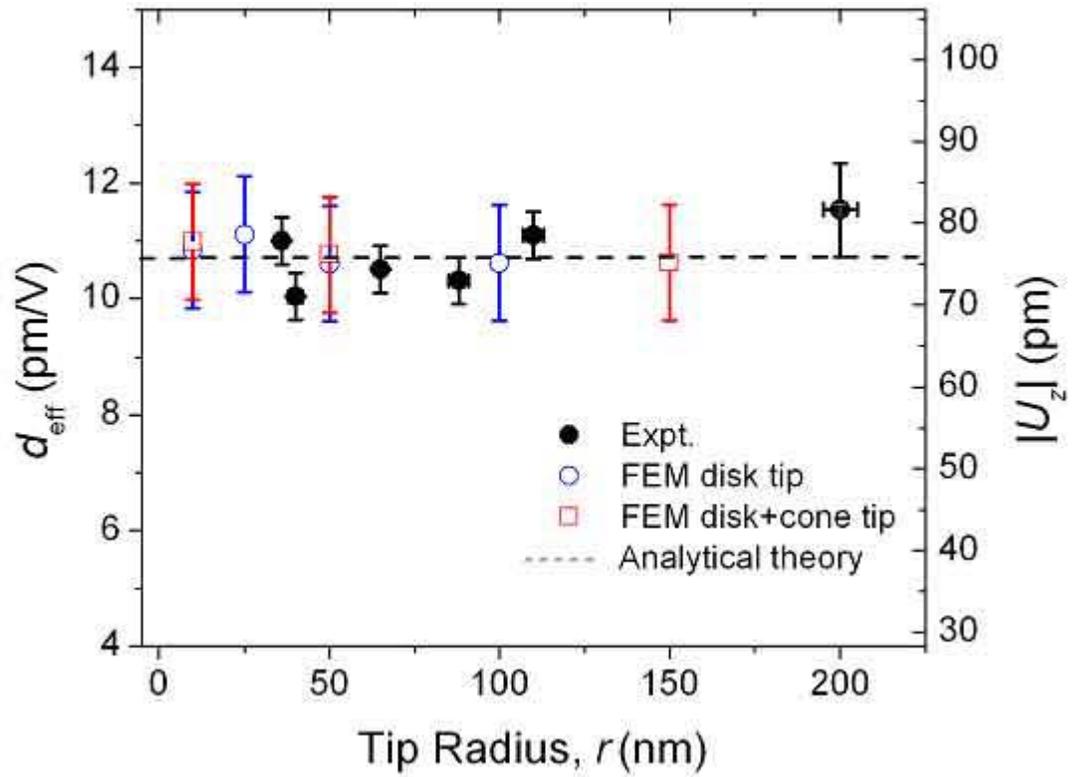

**Figure 3** (color online): The maximum amplitude, $|U_z|$, away from the wall as a function of tip radius, $r$ is shown. Also shown overlapped is the analytical theory and finite element method (FEM) predictions.

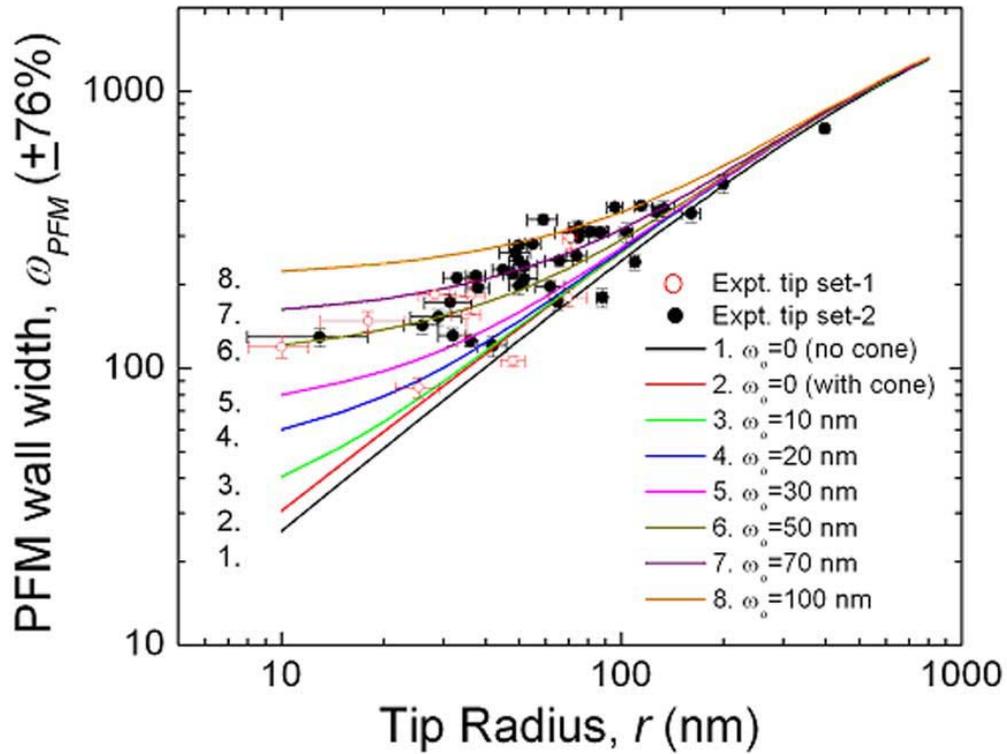

**Figure 4** (color online): PFM wall width as a function of tip radius for sphere-like (tip set-1) and disk-like (tip set-2) PFM tips. Also shown are analytical theory predictions generated from Eqs.(2-5, 8) along with Appendices A and B, for different intrinsic wall half-widths ($\omega_o$). [Reference 47]

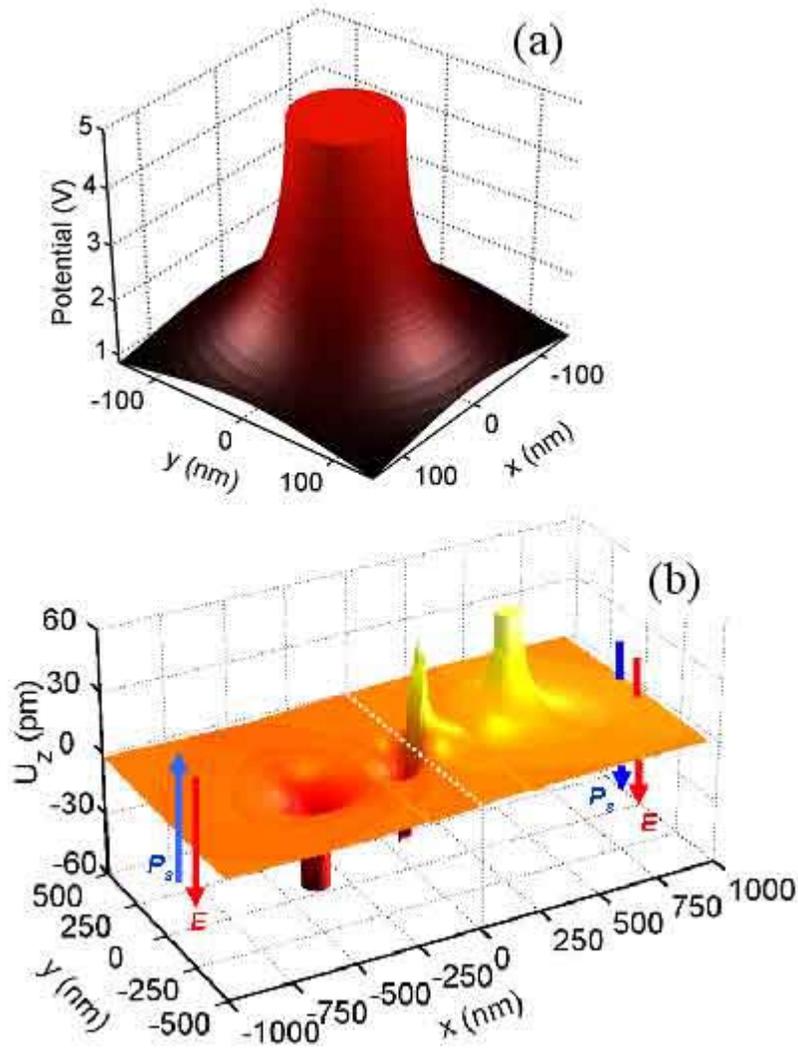

**Figure 5 (color online)** (a) FEM simulated surface potential on LN surface under a 50nm radius disk tip in contact with sample with 5Volts applied. (b) FEM simulated piezoelectric displacements, $U_z$ of a LiNbO$_3$ $z$-surface. Displacements for three different tip locations are shown: tip located on the wall (location 2) and away from the wall on either side (locations 1, 3).

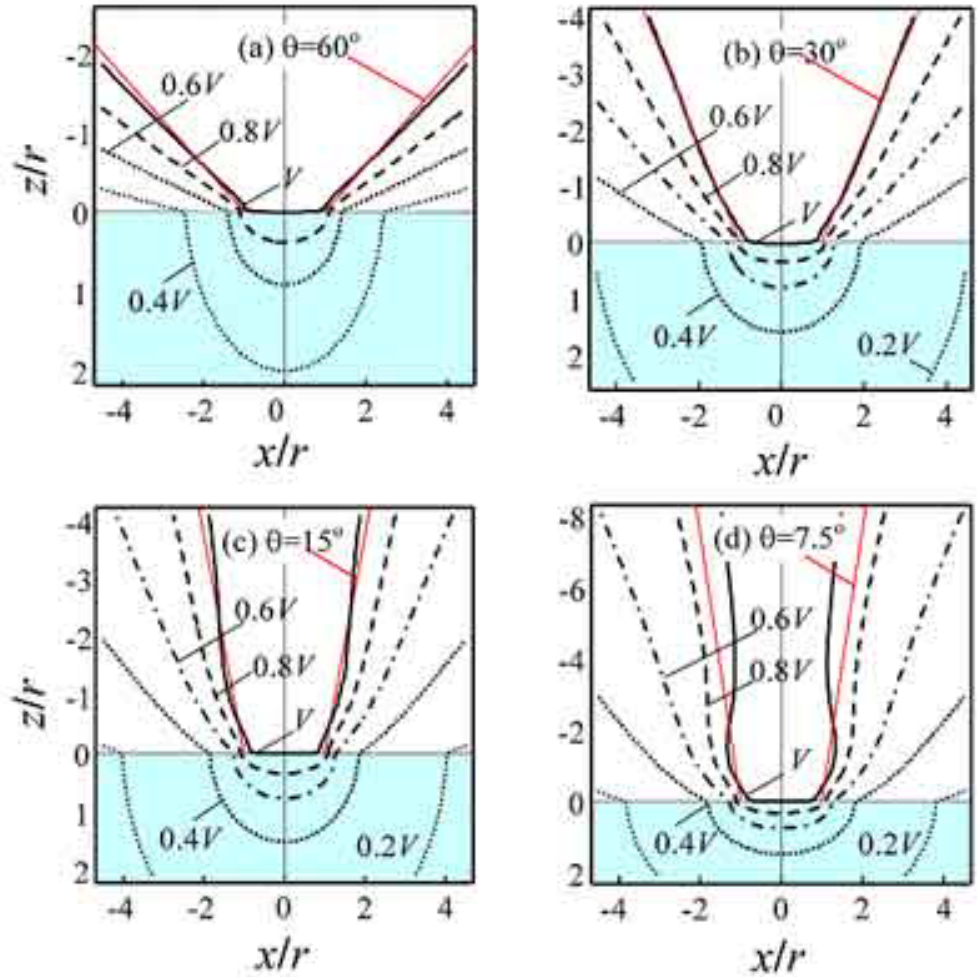

**Figure 6** (Color online) Isopotential lines φ in the vicinity of tip contact with sample surface (boundary air- LiNbO$_3$) for $\Delta L = r$, $q = 4\pi\varepsilon_0 V \Delta L$ and different cone angles $\theta = 60^o$ (a), $\theta = 30^o$ (b), $\theta = 15^o$ (c), $\theta = 7{,}5^o$ (d). Labels near the curves designate potential φ values in applied voltage $V$ units.

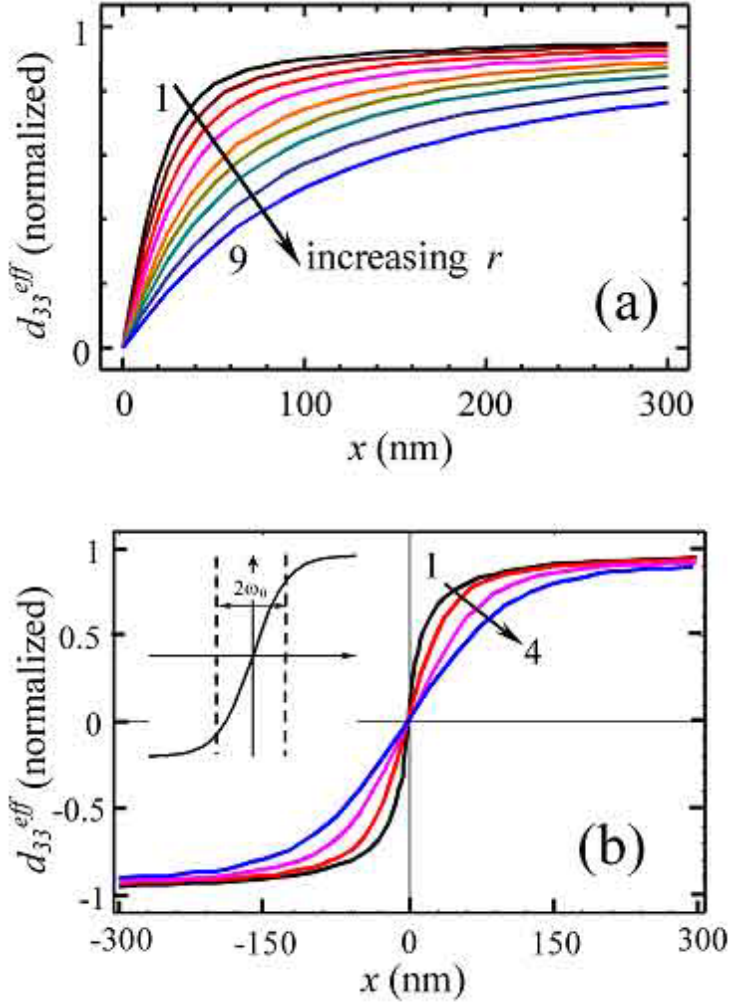

**Figure 7** (Color online) (a). Normalized PFM response profile near a step-like 180° domain wall ($\omega_o=0$) in LiNbO$_3$ for disk radius $r$=17, 25, 36, 50, 70, 88, 110, 150, 200 nm (curves 1-9, respectively), a cone length $L$=20 μm, $\theta = 15^o$ for the PFM tip and an intrinsic halfwidth $\omega_o$=20 nm. (b) Vertical PFM response near the single domain wall in LiNbO$_3$ as a function of the distance from the diffused wall with different halfwidth values $\omega_o$=3, 30, 60, 90 nm (curves 1, 2, 3, 4, respectively); $r$=30 nm, $L$=20 μm, $\theta = 15^o$. The intrinsic wall diffuseness, $d_{lkj}(x) = d_{lkj}^{bulk}\tanh(x/\omega_0)$ is schematically shown in the inset.

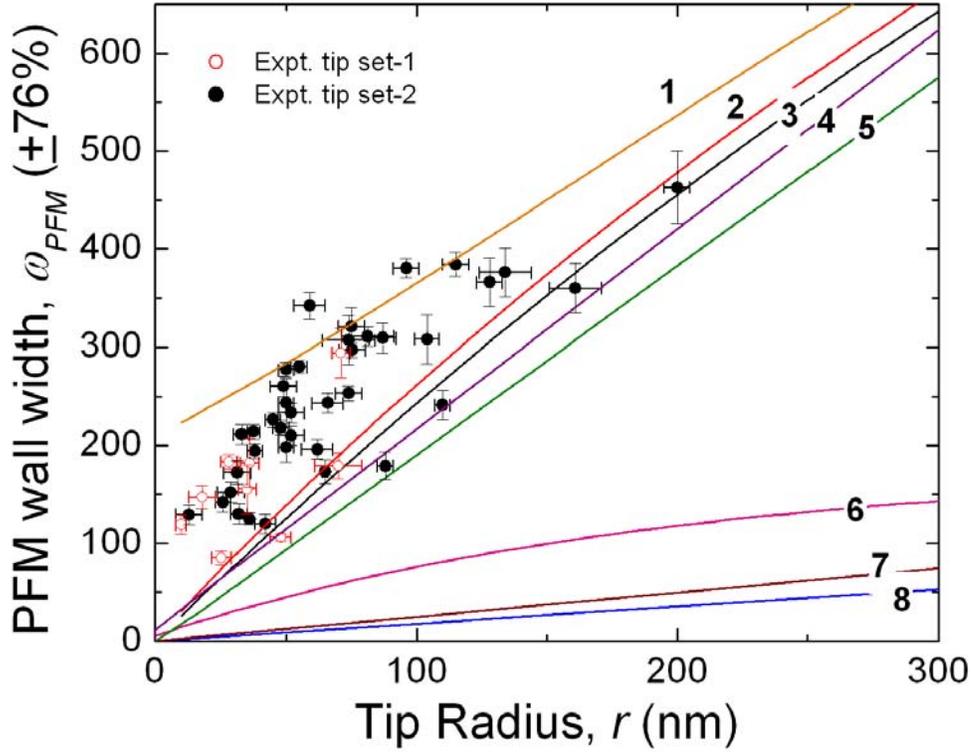

**Figure 8** (Color online): A comparison of different tip models with the experimental data for the $\omega_{PFM}$ across a wall as a function of tip radius, $r$. The experimental data are shown for sphere-type (tip set-1), and disk-type (tip set-2) tips. The different theory curves are: 1) Analytical theory of a diffuse domain wall with $\omega_o$=100nm using a disk-tip including cone section. 2) Analytical theory of a step domain wall with $\omega_o$=0nm using a disk-tip, including cone section. 3) Analytical theory of a step domain wall with $\omega_o$=0nm using a disk-tip, excluding the cone section. 4) FEM with $\omega_o$=0nm using a disk-tip, including the cone section. 5) FEM with $\omega_o$=0nm using a disk-tip, excluding the cone section. 6) FEM with spherical tip (excluding cone) removed at a distance of $d$=2nm from the surface of the sample. 7) FEM with spherical tip including the cone, with $d$=0nm. 8) FEM with spherical tip excluding the cone, with $d$=0nm.

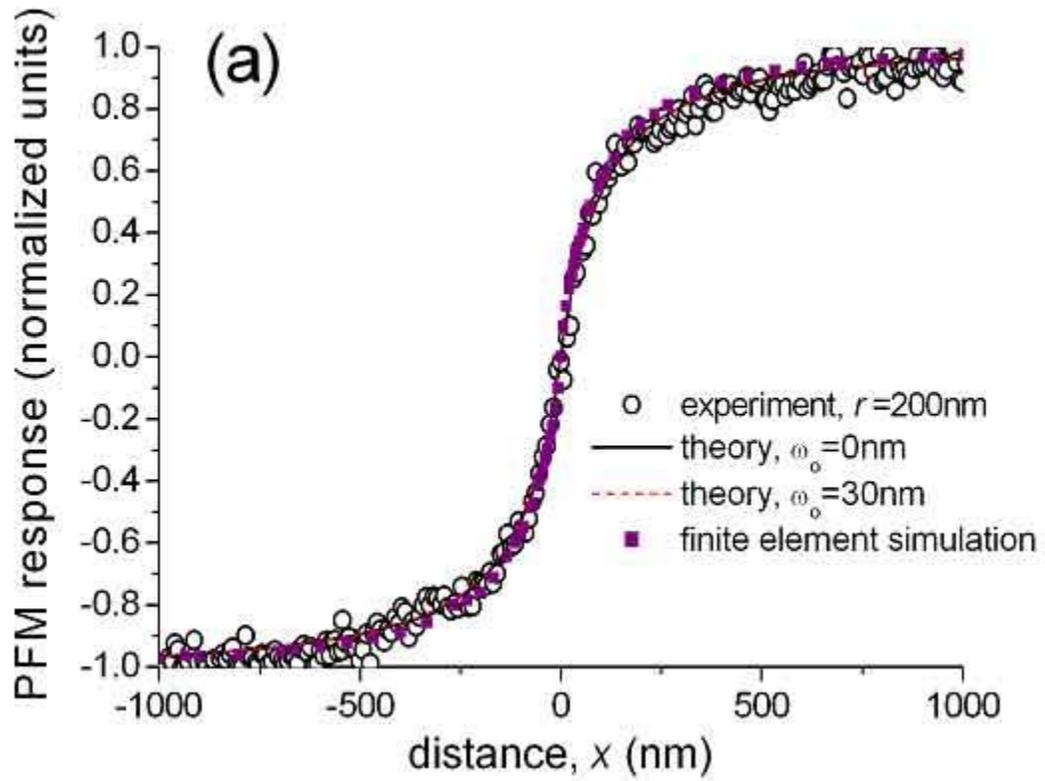

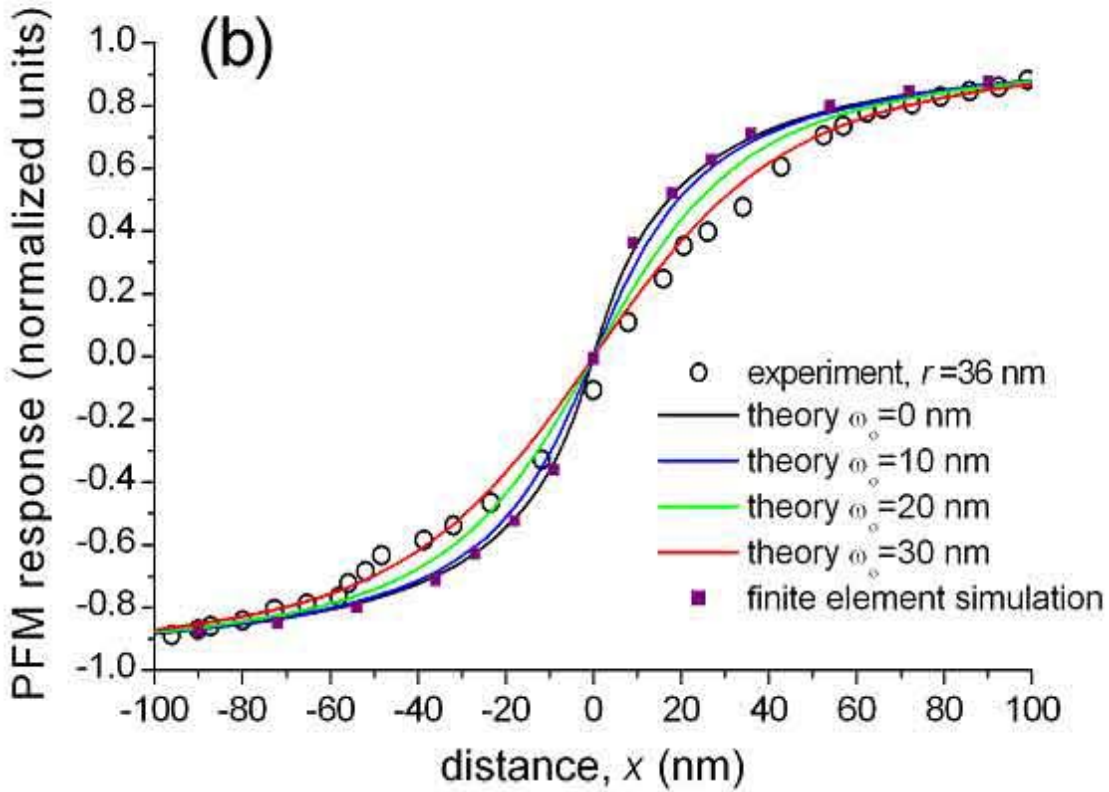

**Figure 9** (color online): Experimental Piezoelectric Force Microscopy (PFM) profile, $U_z$ (open circles) as a function of wall normal coordinate, $x$ across a 180°domain wall in lithium niobate. Measurements were made with a Ti/Pt coated Si-tip with a disk shaped tip actual tip of (a) $r$~36 nm, and (b) $r$=200nm, determined after PFM scan by scanning electron microscopy imaging. An oscillating voltage of 5Vrms, at 42.35kHz was applied to the tip. Theoretical PFM profiles using FEM (solid squares, $2\omega_o$ =0nm) and analytical theory (black, $2\omega_o$ =0nm and red, $2\omega_o$ =30nm) are also shown.

TABLE CAPTIONS:

Table A.1: Relevant physical properties of $LiNbO_3$ and $LiTaO_3$

Table A. 2: Relevant PFM parameters in Eqs.A.1a-d for $LiNbO_3$ and $LiTaO_3$